\documentclass[a4paper,fleqn,usenatbib]{mnras}
\usepackage{mathptmx}
\usepackage[T1]{fontenc}
\usepackage{ae,aecompl}

\usepackage{graphicx}
\usepackage{amsmath}
\usepackage{amssymb}
\usepackage{color}
\newcommand{\kom}[1]{{#1}}

\title[Gamma-ray flares from blobs encountering stars]
{Orphan $\gamma$-ray flares from relativistic blobs encountering luminous stars}
\author[P. Banasi\'nski, W. Bednarek \& J. Sitarek]
{P. Banasi\'nski, W. Bednarek, J. Sitarek \\ 
Department of Astrophysics, The University of \L \'od\'z,
ul. Pomorska 149/153, 90-236 \L \'od\'z, Poland,\\
p.banasinski@uni.lodz.pl, bednar@uni.lodz.pl, jsitarek@uni.lodz.pl}
\begin{document}

\date{Accepted . Received ; in original form }

\pagerange{\pageref{firstpage}--\pageref{lastpage}} \pubyear{2015}

\maketitle

\label{firstpage}

% letter: nie wiecej niz 200 slow w abstrakcie !
\begin{abstract}
We propose that $\gamma$-rays in blazars can be produced during encounters of relativistic blobs of plasma with radiation field produced by luminous stars within (or close to) the jet. The blob is expected to contain relativistic electrons which comptonize stellar radiation to the GeV-TeV energies. Produced $\gamma$-rays can initiate the Inverse Compton $e^\pm$ pair cascade in the stellar radiation. We propose that such a scenario can be responsible for the appearance of the so-called orphan $\gamma$-ray flares. 
We show that the relativistic blob/luminous star collision model can explain the appearance of the extreme orphan $\gamma$-ray flare observed in the GeV and sub-TeV energy range from the flat spectrum radio quasar PKS\,1222+21.
\end{abstract}

\section{Introduction}

It is widely accepted that the non-thermal radiation in blazars is produced within a relativistic jet directed towards the observer. 
However, the mechanism of energization of particles and their radiation is not fully understood. 
Recently, simple scenarios exploiting collisions of compact objects with the jet plasma reached some attention. 
In fact, many different types of compact objects (e.g. stars, clouds or even globular clusters, fragments of supernova remnants or pulsar wind nebulae) can be immersed within the jet, forming an obstacle for the jet plasma. 
It has been proposed that particles, accelerated on the shocks appearing as a result of collisions of the jet plasma with stellar winds, can produce high energy $\gamma$-rays (see e.g. \citealp{bp97,ba10, br12, ar13, wy14, bb15, br15, de16}).
In this work, we consider another encounter scenario in which relativistic particles, already present in the fast moving blob, interact with the dense radiation of a star within the jet. 
Relativistic leptons, which are isotropic in the blob, comptonize stellar radiation to GeV-TeV energies. 
The radiation field of the star is strong enough that $\gamma$-rays produced in the Inverse Compton process can initiate the $e^\pm$ pair cascade in the stellar radiation. To study such a scenario we developed a dedicated Monte Carlo code tracking an electromagnetic cascade in the anisotropic radiation field of a star. 
We calculate the $\gamma$-ray spectra emerging from such process and discuss their features. As an example,
we show that the $\gamma$-ray spectrum and light curve observed from a distant Flat Spectrum Radio Quasar (FSRQ) PKS\,1222+21 during its 2010 orphan $\gamma$-ray flare might be explained in terms of such a scenario.  
 
We propose that blobs of relativistic plasma, moving in jets of active galactic nuclei (AGNs), can encounter from time to time the luminous stars which form a quasi-spherical halo around the central super-massive black hole (see Fig.~\ref{fig1}). 
For example, in the nearby active radio galaxy Cen~A, the current star formation rate ($\sim$2 M$_\odot$ yr$^{-1}$) should result in the production of $(6 -12)\times 10^7$ M$_\odot$ of young stars during the duty cycle of the Cen A active nucleus. 
We can thus expect $\sim$3$\times 10^5$ stars with masses above 20 M$_\odot$ \citep{wy14}. 
A large number of these stars should be immersed in the jet pointing towards the observer.
In the case of a simple conical jet, the number of stars is estimated to be of the order of $\sim$83$\gamma_{30}^{-2}$, where we assumed that the jet opening angle is determined by the Lorentz factor of the flow $\theta \sim 1/\gamma_{\rm b}$ and $\gamma_{\rm b}=30\times \gamma_{30}$. 
In the case of a parabolic jet, the number of stars depends in a more complicated way on the parameters of the jet but is expected to be also significant since the perpendicular extend
of the jet on a parsec distance scale from its base are of similar order (see estimates in Sect.~\ref{sec3}). 
The Cen~A nucleus is also immersed in a quasi-spherical bulge of late type stars with the number estimated as $\sim 8\times 10^8$ \citep{wy14}. 
The total number of luminous red giants can be as large as $\sim$10$^6$, if only $\sim$10$^{-3}$ of these bulge stars are in the red giant phase.
Therefore, the interaction of a relativistic blob with the radiation field of such luminous stars within a jet of an active galaxy seems to be quite likely. 

In Section~\ref{sec3} we explain the basic properties of the blob-star interaction model.
In Section~\ref{sec4} we apply such scenario to the extreme flare observed from PKS\,1222+21. 
\kom{In Section~\ref{sec5}, we summarize our conclusions.}

\section{Gamma-rays from IC pair cascade} \label{sec3}

We assume that the blob, a quasi-cylindrical region within the jet containing relativistic electrons, moves along the jet with a fixed Lorentz factor $\gamma_{\rm b}$. 
The blob is characterised by its radius $R_{\rm b}$ and longitudinal extend, $H_{\rm b}$ (measured in the reference frame of the star).
The radius of the blob has to be smaller than the local perpendicular extend of the jet $R_{\bot}$. In the case of a simple conical jet, $R_{\bot}\approx \theta l = 10^{16}\gamma_{30}^{-1} l_{-1}$ cm, where $l = 0.1l_{-1}$ pc is the distance from the jet base. 
In the case of a parabolic jet, $R_{\bot}$ can be modelled as, $R_{\bot} = al^{1/2} + R_{\rm s}\approx 2\times 10^{15}(1 + 22l_{-1}^{1/2})$\,cm.
As an example, the parameter $a$ is obtained from the observation of the jet in the nearby radio galaxy M87 (see Sect.~4.4 in \citealp{an12}), 
and $R_{s} = 2\times 10^{15}$ cm is the Schwarzschild radius of the M87 black hole.
We assume that the longitudinal extend of relativistic electrons within the blob can be described by a Gaussian distribution with a standard deviation of $H_{\rm b}$. Such a blob, when passing the stellar radiation, assumed to be point like, should produce a flare on the time scale determined by $H_{\rm b}$. % $\tau_{\rm f}$
Note that the emission is produced by successive layers of the blob as they enter in the fixed volume around the star, rather than from the whole blob traversing the jet.
Thus, the observed duration of the emission is not going to be shortened by large values of $\gamma_b$.

\begin{figure}
\centering
\includegraphics[width=0.35\textwidth]{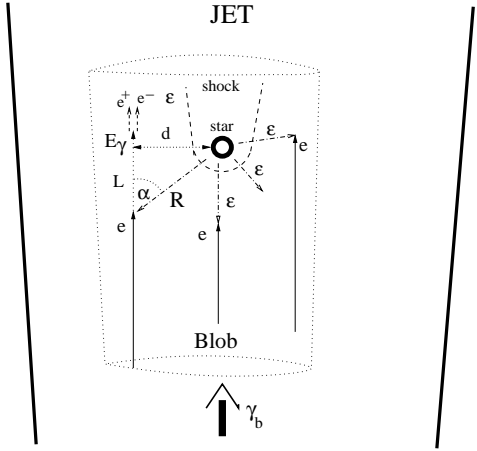}
\caption{Schematic representation of the interaction of a cylindrically shaped blob with the radiation of a luminous star. The blob, containing relativistic electrons (isotropic in the blob reference frame), moves within the jet with the Lorentz factor $\gamma_{\rm b}$.  
A specific electron in the blob moves with some impact parameter, $d$, in respect to the star, at the distance $L$ measured along the jet. Electrons produce $\gamma$-rays in the Inverse Compton process which can be farther absorbed in these same stellar radiation. Secondary e$^\pm$ pairs can interact with the stellar photons ($\varepsilon$) initiating the IC $e^\pm$ pair cascade.}
\label{fig1}
\end{figure}
%

%
%
%\section{Gamma-rays from IC pair cascade} \label{sec3}

Relativistic electrons in the blob can suffer strong energy losses on comptonization of radiation coming from a single star provided that the star is close to the direction of a moving blob. 
We assume that electrons are already injected into the blob with a power law spectrum. 
They are isotropic in the blob reference frame and the blob moves relativistically through the jet. 
The acceleration mechanism of electrons and their maximum energies are not specified in our model. 
They depend on the acceleration efficiency of electrons and of their dominant energy loss process. 
We assume that electrons can reach at least sub-TeV energies which in the case of the acceleration process saturated by the synchrotron energy losses would require the magnetic field within the blob below several Gauss (see e.g. Eq.~3 in \citealp{sb10}). 
Electrons with sub-TeV energies in the relativistic blob can scatter the stellar radiation already quite deep in the Klein-Nishina (KN) regime. Therefore, $\gamma$-ray photons produced in the IC process can have energies comparable to energies of the primary electrons. 
These $\gamma$-rays can be again absorbed in this same radiation field, initiating an IC $e^\pm$ pair cascade in the radiation of the luminous star. 

\begin{figure}
\includegraphics[width=0.235\textwidth]{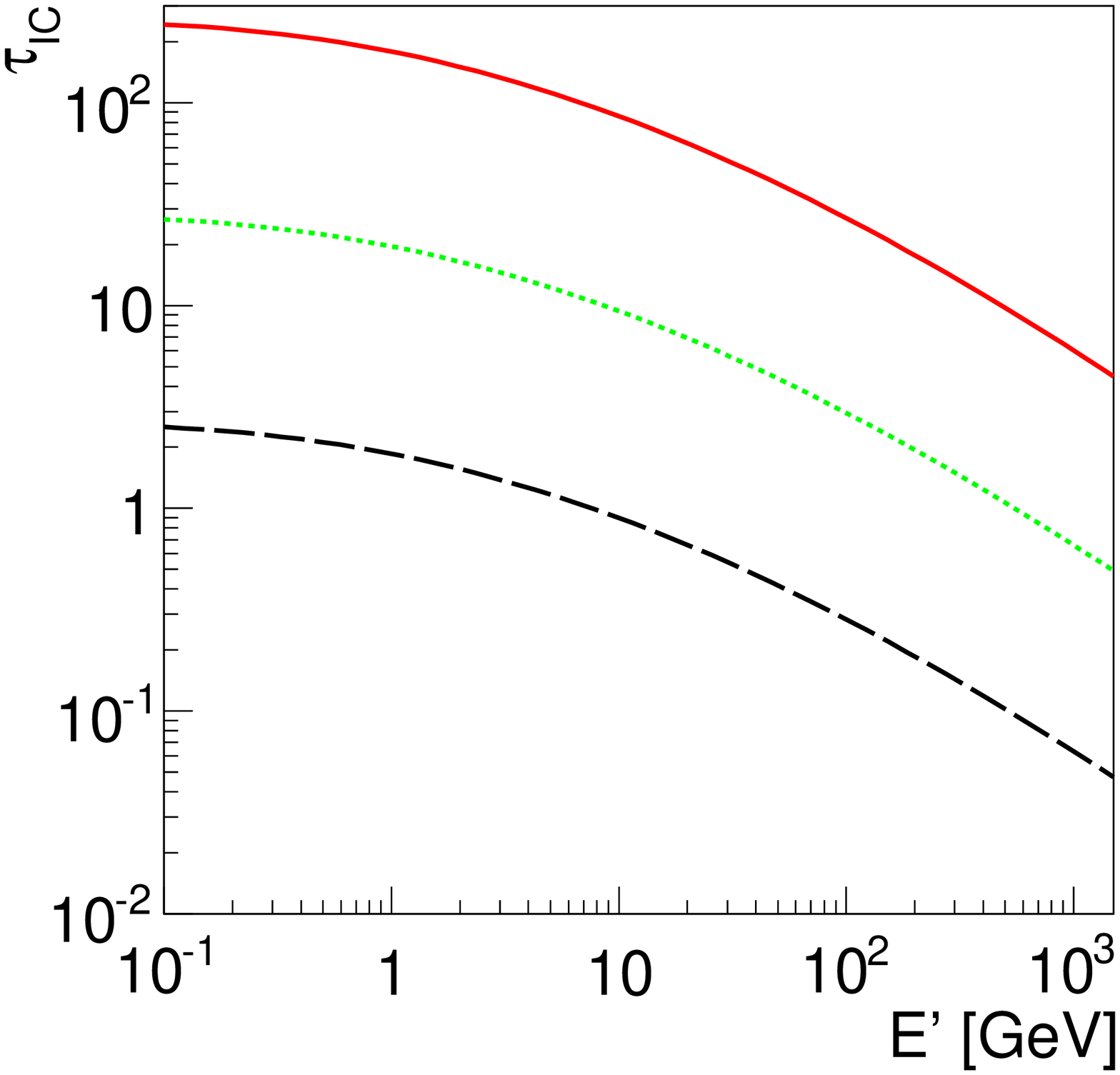}
\includegraphics[width=0.235\textwidth]{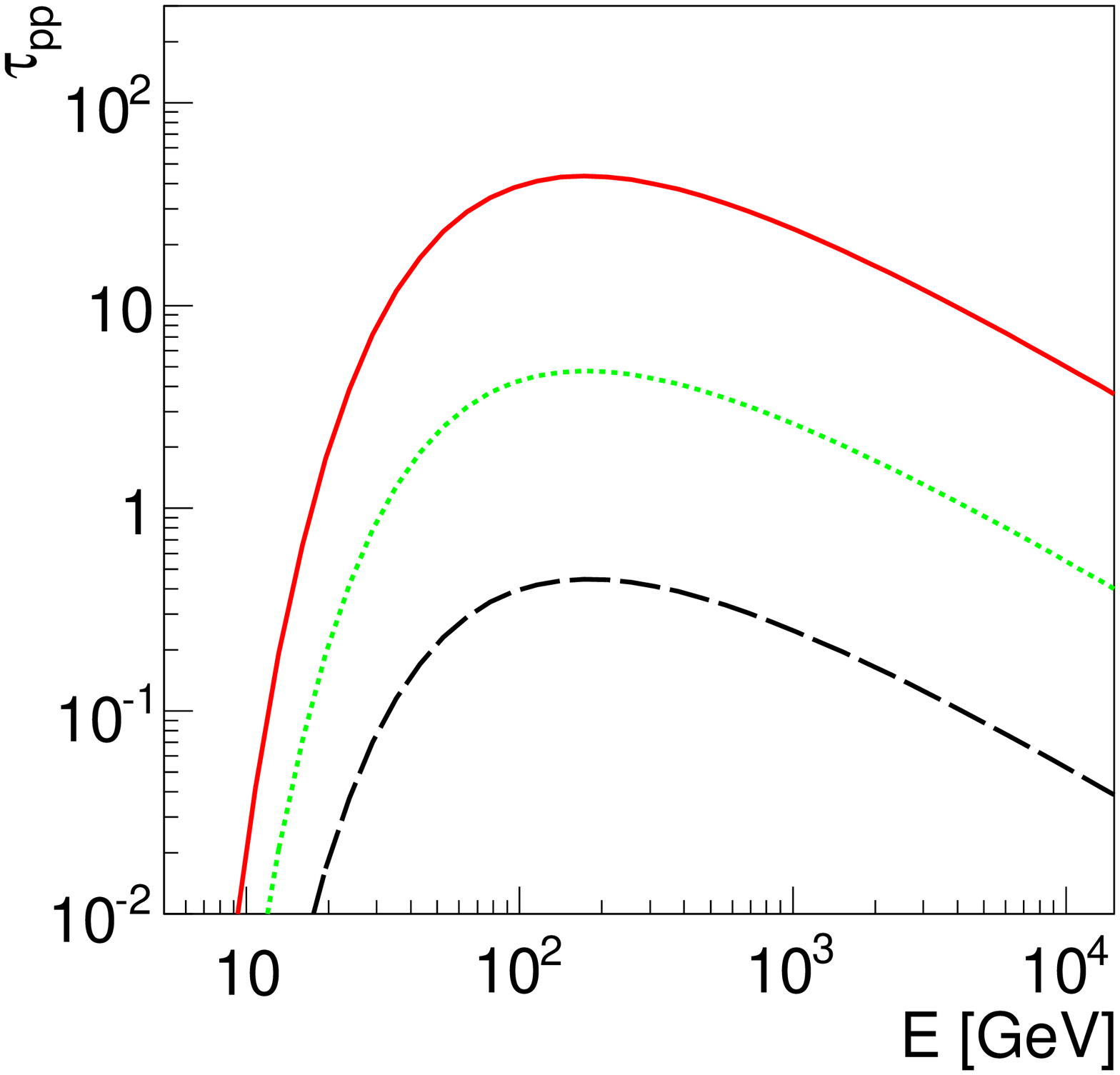}
\caption{
Optical depth for the inverse Compton scattering of the stellar radiation by electrons in the blob (the left panel) and for absorption of produced $\gamma$-rays in this stellar radiation (the right panel) as the function of the energy (measured in the reference frame of the blob with $\gamma_b=10$ for the electrons and in the reference frame of the star for the $\gamma$-rays). The impact distance, $d$, is equal to $1.1\,R_\star$ (red, solid lines), $10\,R_\star$ (green dotted) and $100\,R_\star$ (black dashed). 
}
\label{fig2}
\end{figure}

Let us consider a single electron at a specific distance, $L$, from the perpendicular plane containing the star and with the impact parameter $d$ (see Fig.~\ref{fig1}). The direction of the electron is isotropised in the blob by the random magnetic field.  
In order to determine whether the electron can produce a $\gamma$-ray photon and whether the photon can be absorbed in the stellar radiation, we calculate their IC scattering mean free path and the $\gamma-\gamma \rightarrow e^\pm$ mean free path for specific impact distances (see Fig.~\ref{fig2}).
In the left panel we show the optical depth for the Inverse Compton (IC) process of electrons on the stellar radiation as the function of their energy measured in the blob's frame of reference. 
In the calculations, as an example, we assume an O type star with a temperature $T_\star = 3\times 10^4$ K and a radius of $R_\star = 10^{12}$\,cm. 
The optical depth is inversely proportional to $d$. 
The electrons can efficiently produce $\gamma$-rays up to a distance $d\sim 10^{14}$\,cm. 
At the highest energies the optical depth is diminished by the KN effect.
The right panel of Fig.~\ref{fig2} shows the optical depth for the absorption of $\gamma$-rays in the $e^\pm$ pair production process. For the assumed temperature of the star the peak of the pair production cross section falls around 100\,GeV. Large values of the optical depth lead to a strong absorption of such $\gamma$ rays passing within the region of the inner few tens of stellar radii around the star.
Some photons might however undergo a few generations of an electromagnetic cascade and thus still be able to escape this region.

\begin{figure}
\centering
\includegraphics[trim={0 48 0 0 },  clip=true, height=3.75cm]{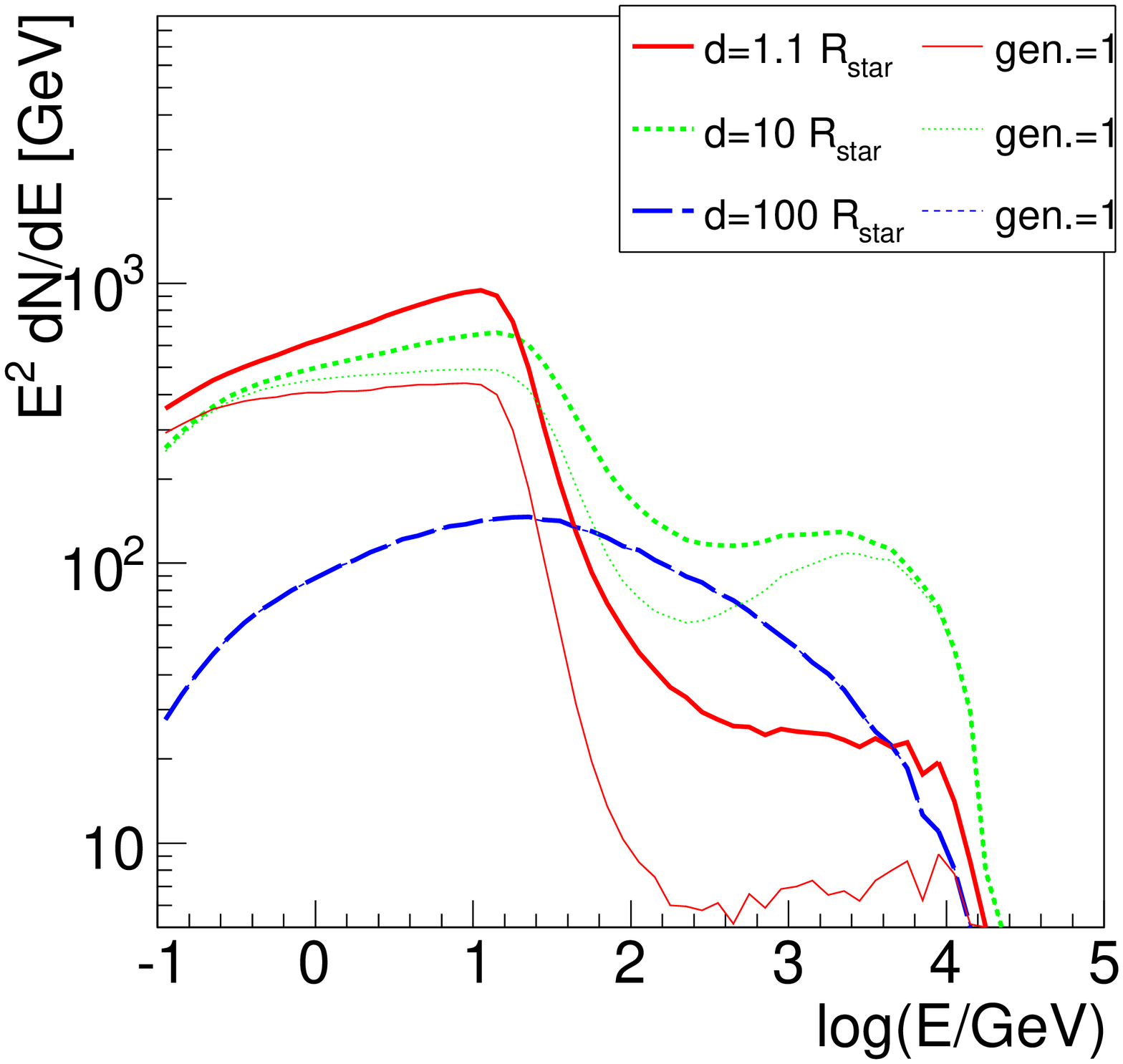}
\includegraphics[trim={50 48 0 0 }, clip=true, height=3.75cm]{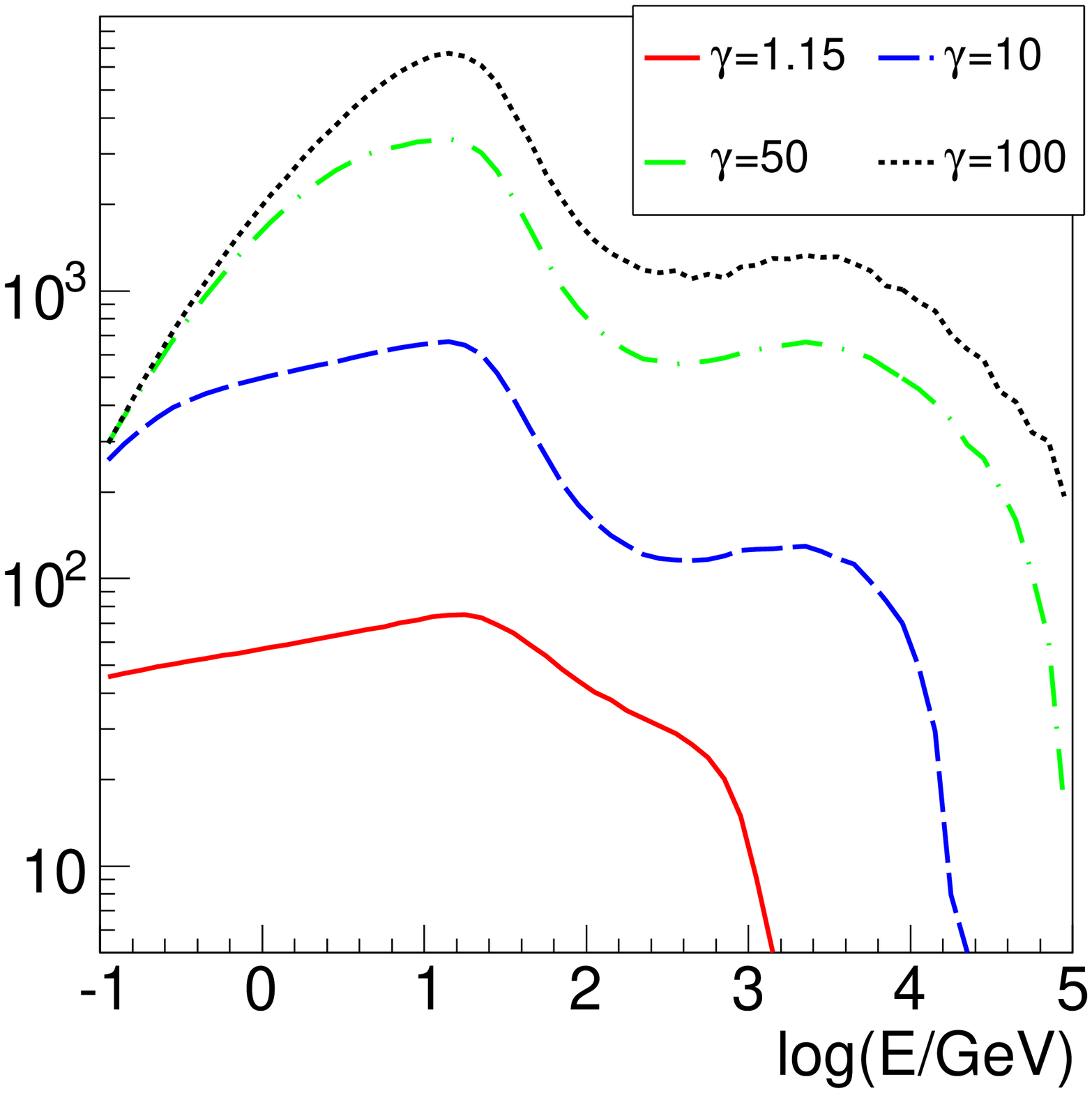}\\
\includegraphics[trim={0  0 0 0 }, clip=true, height=4.1cm]{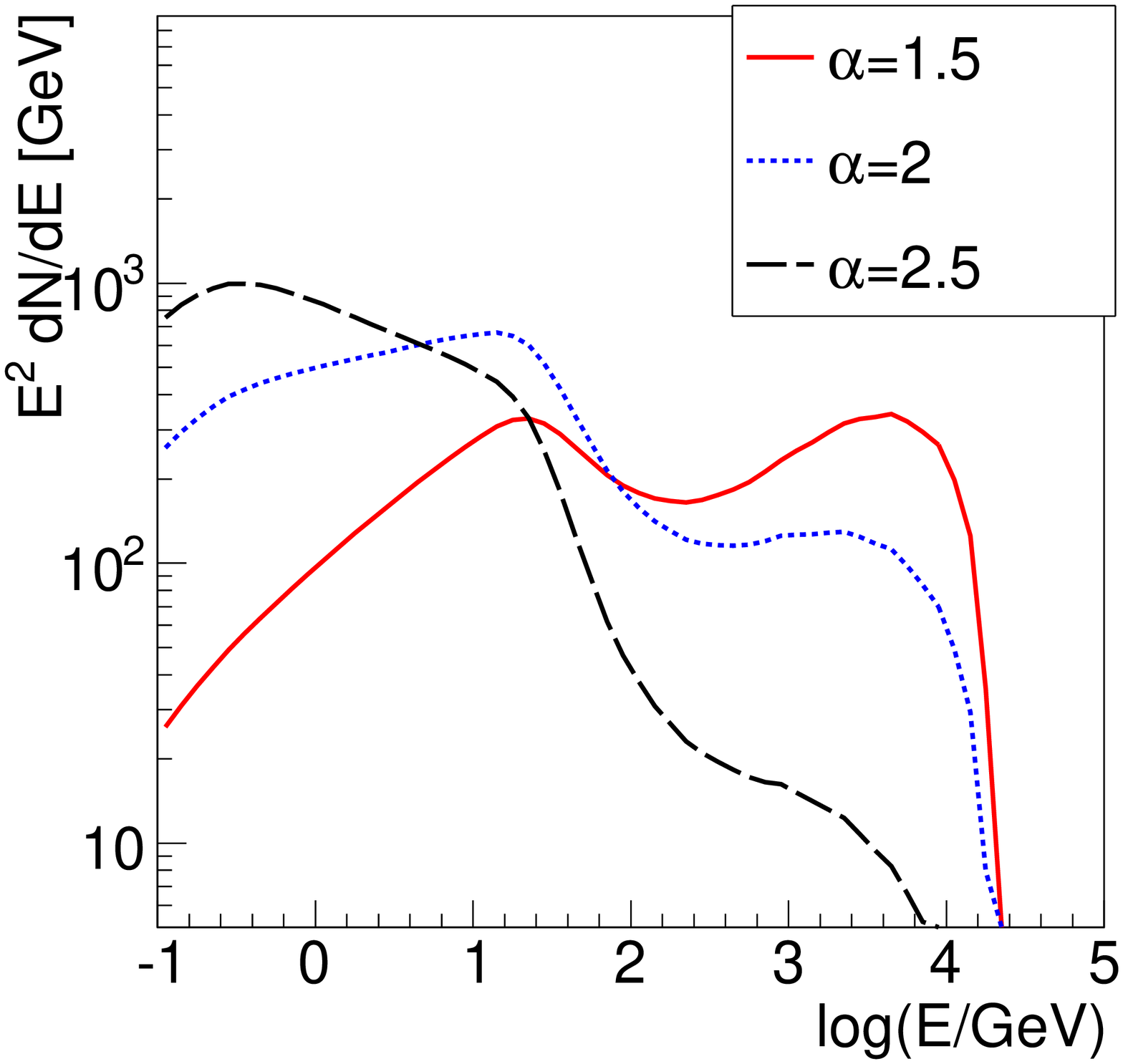}
\includegraphics[trim={50 0 0 0 }, clip=true, height=4.1cm]{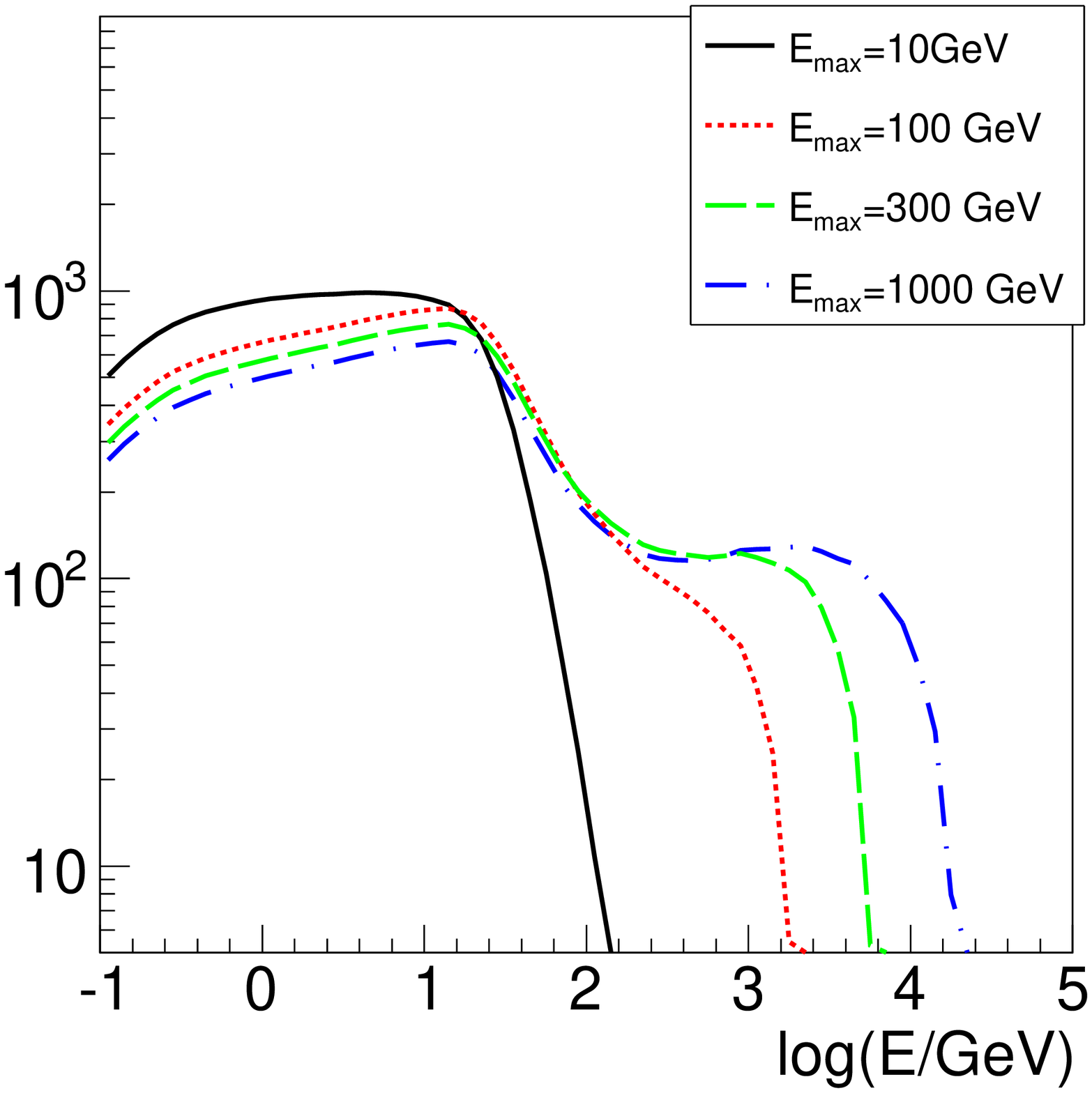}
\caption{
Spectral energy distribution (SED) of the $\gamma$-ray spectrum for a power law differential spectrum of electrons with an spectral index of $\alpha$ between 0.1 GeV and $E'_{\max}$ for fixed impact distance of electrons. 
Top left panel: dependence on impact distance $d=1.1\,R_\star$ (red solid), $10\,R_\star$ (green dotted), $100\,R_\star$ (blue dashed), thick lines show the spectra from the full cascade, thin lines show the spectra escaping from the first generation of photons.
Top right panel: dependence on the Lorentz factor of the blob $\gamma_b=$ 1.15 (red solid), 10 (blue dashed), 50 (green dot-dashed) and 100 (black dotted). 
Bottom left panel: dependence on the spectral index of electrons 
$\alpha=1.5$ (red solid), 2 (blue dotted) 2.5 (black dashed).
Bottom right panel: dependence on the maximum energy of the electrons (measured in the blob's frame): $E'_{\max}=10\,$GeV (black solid), 100\,GeV (red dotted), 300\,GeV (green dashed), 1000\,GeV (blue dot-dashed).
Unless specified other\-wise $\gamma_b=10$, $d=10\,R_\star$, $\alpha=2$, $E'_{\max}=1\,$TeV.
The spectra are normalised to 1\,erg of injected electron energy in the blob's frame of reference. 
}
\label{fig3}
\end{figure}

In Fig.~\ref{fig3} we study the dependence of the observed emission on various parameters of the relativistic electrons. For simplicity we normalise the escaping $\gamma$-ray emission to 1\,erg of energy injected in the blob (measured in its frame).  The emission is integrated over the whole range of the observation angles. 
In the top left panel, we investigate how the $\gamma$-ray spectrum depends on the impact parameter of the primary electrons. For electrons passing close to the star, the $\gamma$-ray absorption in the stellar radiation makes a clear imprint on the spectrum. Interestingly, even while the optical depths for TeV photons passing next to the star reach values $\gtrsim10$, there is still a measurable flux of TeV photons emitted for electrons passing that close to the star surface.  It is a combined effect of the cascading process and the fact that the impact parameters of electrons are usually smaller than those of secondary $\gamma$-rays. 
To explain the second effect, let us consider a $\gamma$ ray produced by an electron in a blob approaching a star ($d\approx0$, see Fig.~\ref{fig1}) at the distance $L$.
Due to the anisotropy of the electron direction the $\gamma$ ray is expected to be produced at an angle $\theta_\gamma\sim1/\gamma_{\rm b}$ with respect to the blob movement direction. 
Thus, the impact parameter of the $\gamma$ ray, $\sim L \theta_\gamma$, might be much larger than the impact parameter of its parent electron, provided that the $\gamma$ ray was emitted far from the star.
Moreover, if the optical depth is of the order of $10$, the $\gamma$-ray might still escape the production region with similar energy, undergoing a few generations of an electromagnetic cascade if the IC process occurs in KN regime (compare the thin and thick red curve in the top left panel of Fig.~\ref{fig3}). 
The cascading effects becomes negligible for the impact factor of an electron clearly exceeding the value of $\sim 10R_\star$.

In the top right panel in Fig.~\ref{fig3}, we investigate the dependence of the $\gamma$-ray spectra on the Lorentz factor of the blob. 
As the electron energy is normalised to a value calculated in the blob's frame of reference, for large values of $\gamma_{\rm b}$ the emission is magnified by $\sim\gamma_{\rm b}$ and extends to higher energies. 
As expected, the spectral shape also depends strongly on the spectral index, $\alpha$, of the power law spectrum of electrons (see the bottom left panel of Fig.~\ref{fig3}). 
For very flat electron spectra, the visibility of the absorption dip is influenced by the maximum energies of electrons (see the bottom right panel of Fig.~\ref{fig3}). 

\begin{figure}
\includegraphics[width=0.235\textwidth]{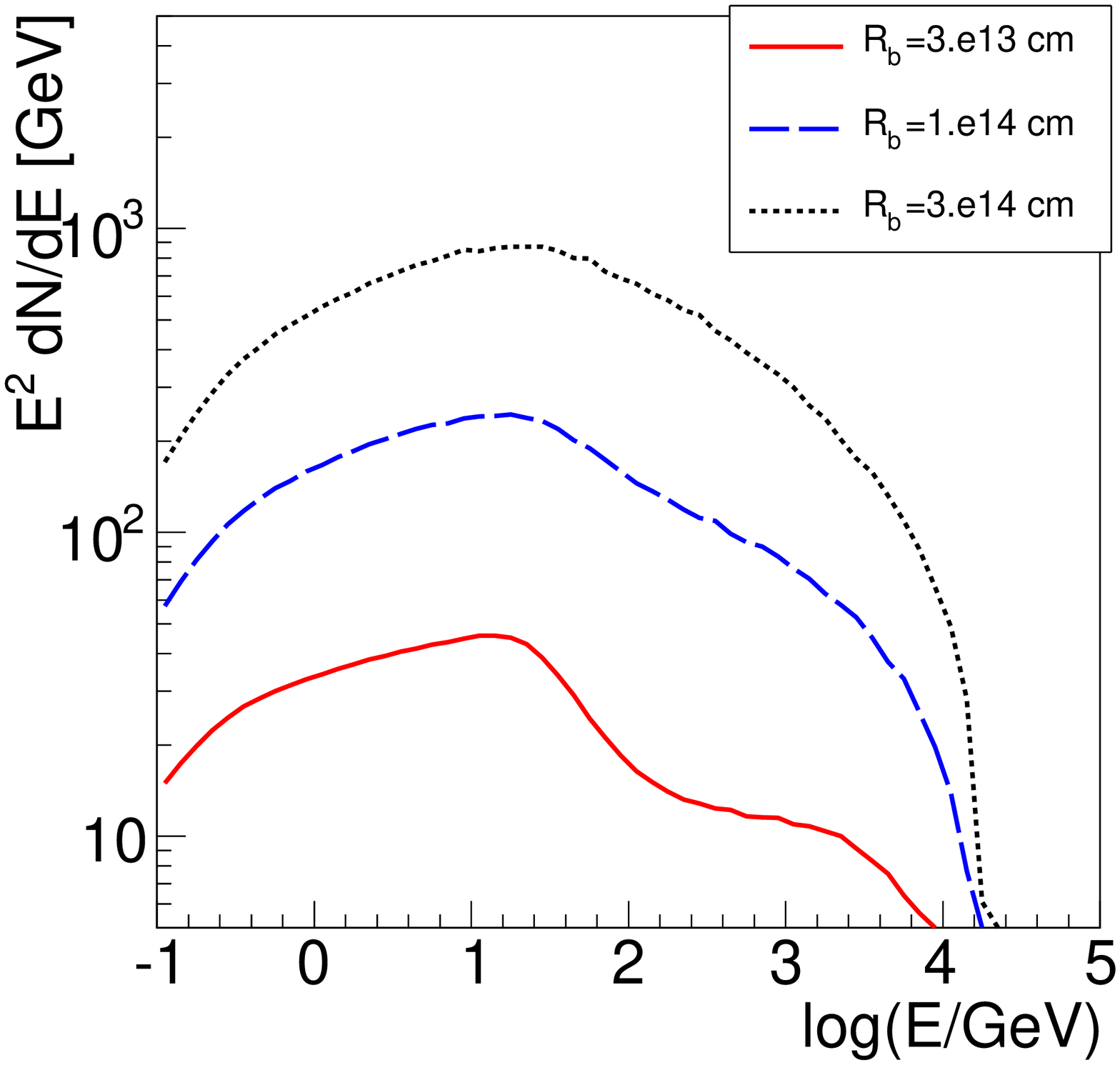}
\includegraphics[width=0.235\textwidth]{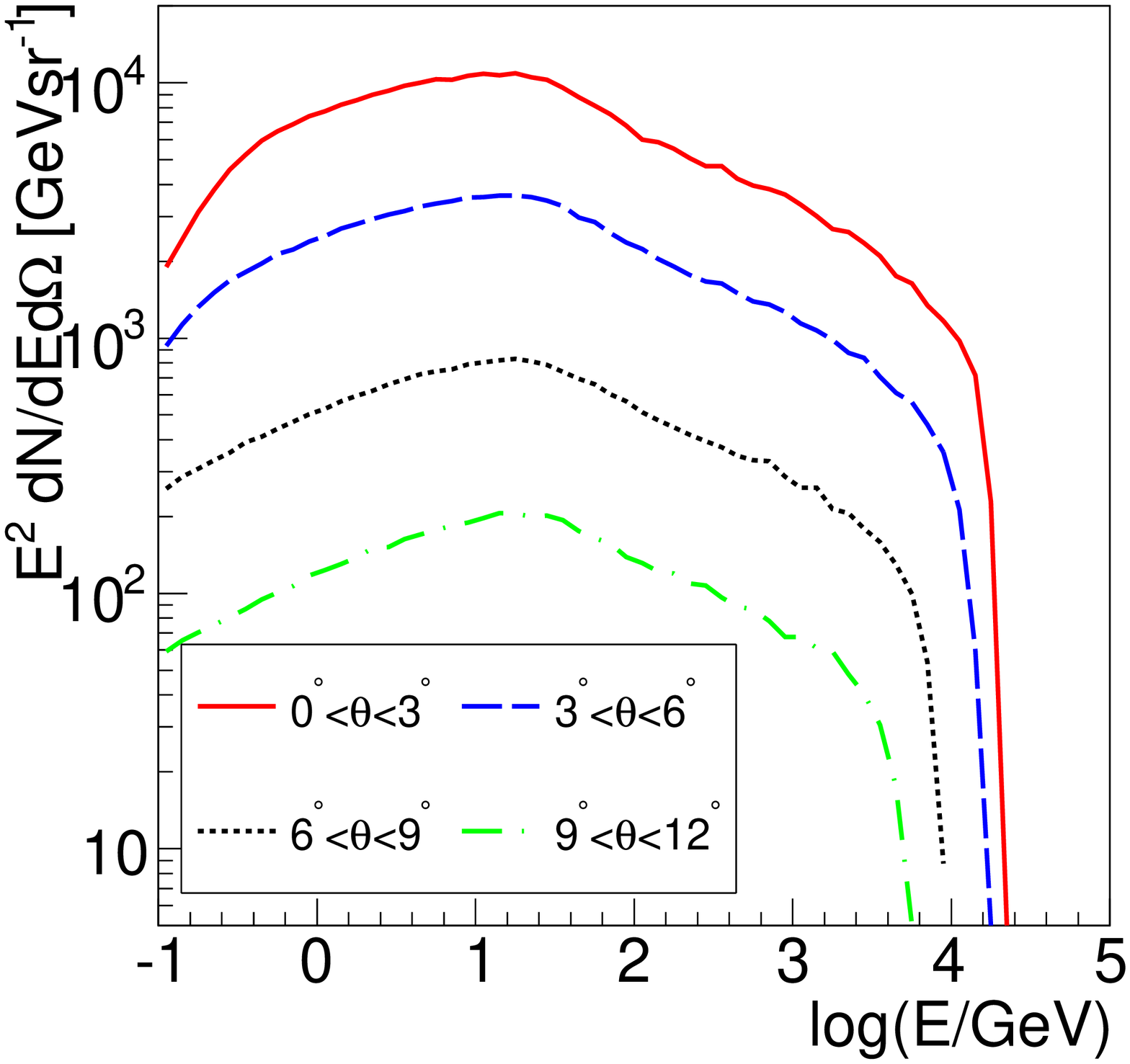}
\caption{
SED from a cylindrical blob filled homogeneously with electrons with a power law energy distribution between 0.1 and 1000 GeV and a spectral index of 2. 
The total energy in the electron spectrum is normalised to 1\,erg per $\pi (10^{14}\,\mathrm{cm})^2$ cross section of the blob. The blob is moving with $\gamma_b=10$.
Left panel: $\gamma$-ray spectra for different radii of the blob: $3\times10^{13}$\,cm (red, solid), $10^{14}$\,cm (blue, dashed), $3\times10^{14}$\,cm (black, dotted). 
Right panel: spectra emitted within the solid angle observed at different range of observation angles $0^\circ-3^\circ$ (red solid), $3^\circ-6^\circ$ (blue dashed), $6^\circ-9^\circ$ (black dotted), $9^\circ-12^\circ$ (green dot-dashed). The radius of the blob is equal to $10^{14}$\,cm. 
}
\label{fig4}
\end{figure}

As the next step, we calculate the expected emission from an extended cylindrical blob with a given radius. 
We show the dependence of the obtained spectrum on the radius of the blob (the left panel of Fig.~\ref{fig4}). 
The spectra are normalised to a constant column energy density, namely the total energy of the electrons measured in the blob's frame is 1\,erg per $\pi (10^{14}\,\mathrm{cm})^2$ cross section of the blob. 
The absorption dip is clearly visible for the blobs with the radii $R_{\rm b} \lesssim 30\,R_\star$. 
As long as the IC scattering on the stellar radiation is the dominant energy loss process for the electrons, the emission roughly scales with the radius of the blob cylinder. 
This is the effect of the inverse linear dependence of the IC optical depth on the impact parameter of the electrons (see Fig.~\ref{fig2}). 
As the blob originates far from the star, and no other energy loss processes are taken into account, the dependence of the obtained spectrum on the longitudinal extend of the blob is trivial (scaling with the total amount of energy available in the blob).
Note that due to beaming effect of the blob (anisotropy of the electron directions and their energies in the reference frame of the star), the $\gamma$-ray spectra are strongly dependent on the observation angle.
They reach higher fluxes and larger energies for the observation angle closer to the jet axis (see the right panel of Fig.~\ref{fig4}).

\section{Interpretation of PKS\,1222+21}\label{sec4}

PKS\,1222+21 (4C+21.35) is a flat spectrum radio quasar (FSRQ) at the distance corresponding to the redshift $z = 0.432$ \citep{op87}. The jet moves at a small angle to the observer's line of sight with the apparent superluminal motion $\beta_{\rm app} > 10$ \citep{ho92,jo01,ho01}. 
PKS\,1222+21 has been detected by {\it Fermi}-LAT in the GeV energy range \citep{lo09,ci09} and also by AGILE \citep{ia10}. First two years of the {\it Fermi}-LAT observations of PKS\,1222+21 has been studied in detail in \citet{ta11}. Two large outbursts have been observed from this source in April and June 2010. 
The $\gamma$-ray spectrum is well described by the power law with the differential spectral index close to $-2$ below 1-3 GeV with a clear steepening above this energy (spectral index in the range $\sim$2.4-2.8, \citealp{ta11}). 
During the second {\it Fermi}-LAT detected flare the source was also discovered by the MAGIC Collaboration in the energy range 70-400 GeV \citep{al11}. The observed MAGIC spectrum is well fitted by the power law with a spectral index $3.75\pm 0.27_{\rm stat}\pm 0.2_{\rm syst}$. 
The sub-TeV emission has shown a rapid variability with the doubling time of the order of $\sim$10\,min \citep{al11}.  
A rather weak synchrotron emission of PKS\,1222+21 is partially obscured by a stronger emission from the accretion disk and the dusty torus (e.g. \citealp{ac14}). 
The isotropic power emitted in $\gamma$-rays is $L_\gamma\cong 7\times 10^{47}$ erg s$^{-1}$ (\citealp{ta11}). The mass of the black hole within PKS\,1222+21 is expected to be in the range $6-8\times 10^8$ M$_\odot$ \citep{fa12,ac14}. The above described emission features are difficult to understand in terms of any proposed models for the $\gamma$-ray production (see e.g. discussion in~\citealp{ta11, al11, ac14}).

Here we propose that the $\gamma$-ray flare can be interpreted in terms of the model discussed in Section~\ref{sec3}, i.e. as a result of a collision of the relativistic blob in the jet with a luminous star. 
In fact the low energy spectrum, in the X-rays and the part below UV, does not change significantly between low and high $\gamma$-ray states (see e.g. Fig.~9 in \citealp{ac14}). But the GeV-TeV $\gamma$-ray emission changes by over an order of magnitude. 
Such spectral behaviour resembles emission observed during the so called orphan $\gamma$-rays flares reported earlier from e.g. BL Lac type active galaxy 1ES 1959+650 (\citealp{kr04}). 

\begin{figure}
\centering
\includegraphics[width=0.43\textwidth]{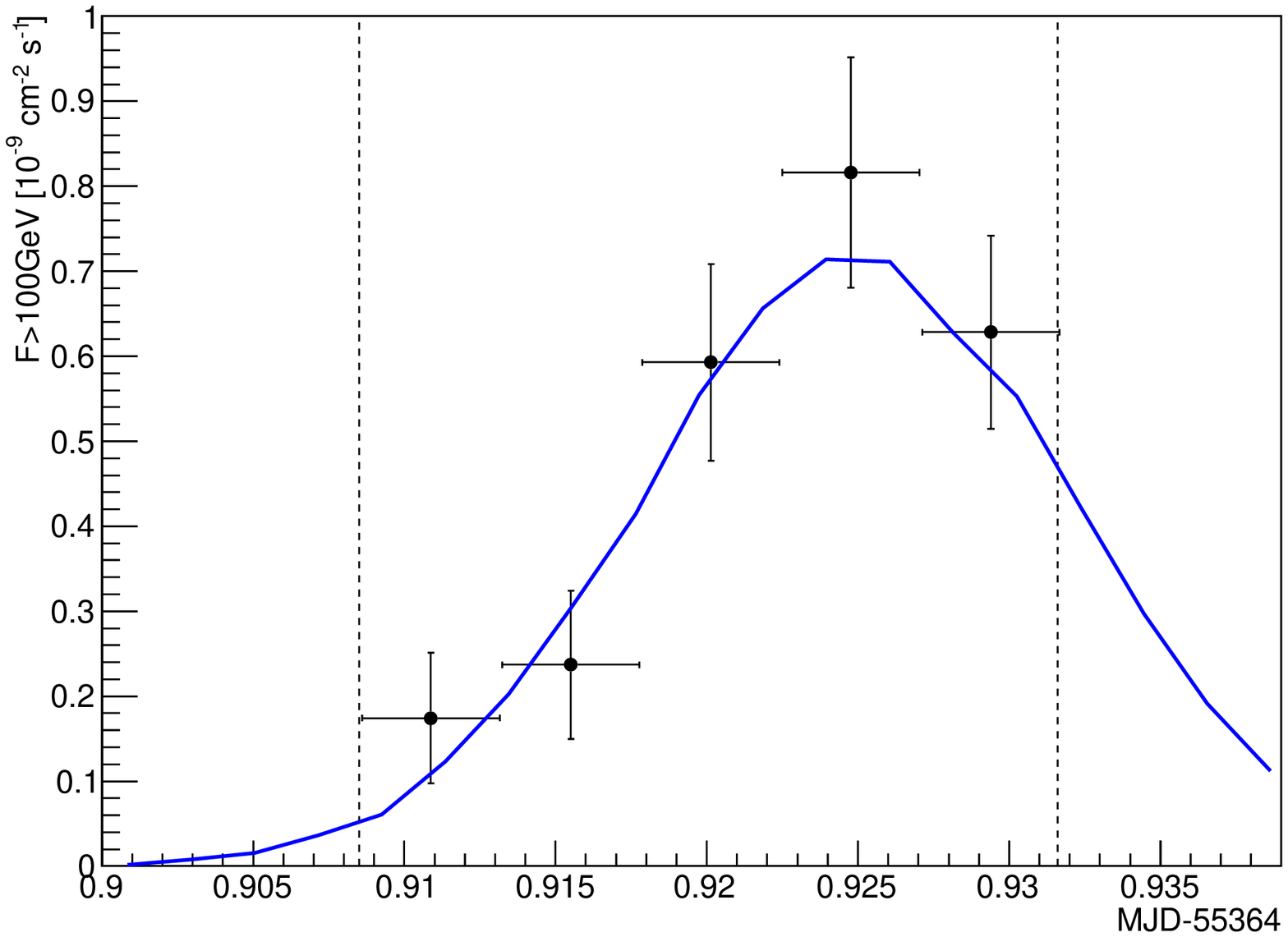}
\includegraphics[width=0.43\textwidth]{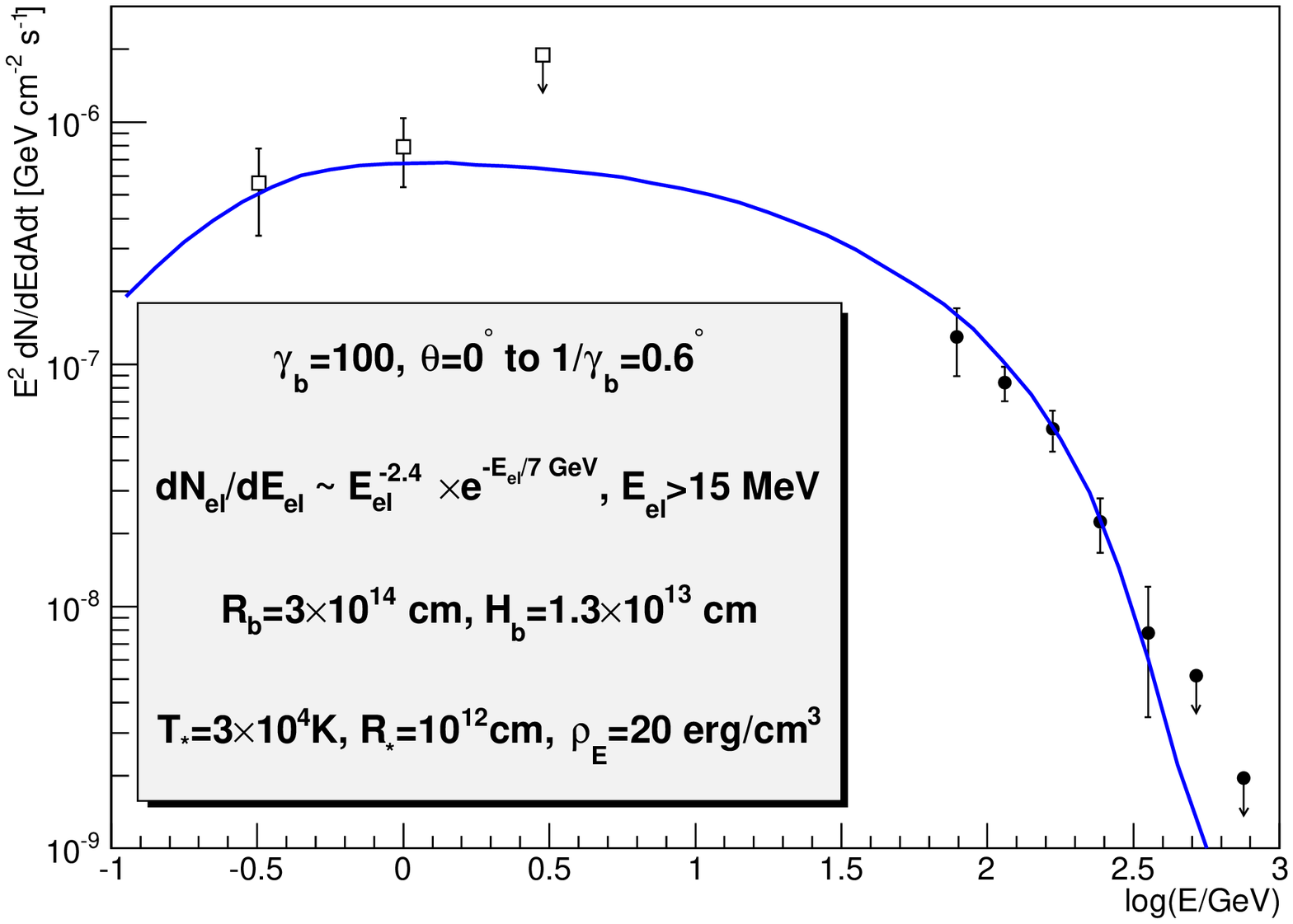}
\caption{
Interpretation of the $\gamma$-ray emission observed during the flare from the FSRQ PKS\,1222+21 in June 2010 by \textit{Fermi}-LAT (empty squares) and MAGIC (full circles), \citet{al11}.
The absorption in the Extragalactic Background Light is taken into account according to \citet{do11} model. 
The top panel shows the light curve above 100 GeV.
The dashed vertical lines are the time range from which the spectral energy distribution (the bottom panel) is computed.
The parameters of the blob, star and electron distribution are given in the inset tab.
}
\label{fig5}
\end{figure}

As an example, we interpret the $\gamma$-ray emission from the flare observed by the {\it Fermi}-LAT and MAGIC as a result of an encounter of a luminous O type star by a relativistic blob in the jet.  
The time scale of the flare observed by the MAGIC telescopes puts a strong constraint on the longitudinal shape and extend of the blob. We assume in the calculations a Gaussian distribution of the electrons with a standard deviation (measured in the reference frame of the star) of $H_{b}=13\,R_\star$.
Such a shape results in a $\gamma$-ray light curve consistent with the one observed by the MAGIC telescopes (see the top panel of Fig.~\ref{fig5}). 

We calculate the corresponding $\gamma$-ray spectrum constructed from the photons arriving at the observer within the observation time of MAGIC and plot it in the bottom panel of Fig.~\ref{fig5}.
Although it is possible to explain the very fast variability of the emission even with a moderate Lorentz factor of the blob, strong constraints are put by the level of the observed flux. 
If the emission region has a radius of $R_{\rm b}=3\times 10^{14}$\,cm, and is moving with a Lorentz factor of $\gamma_{\rm b}=100$, we require an energy density of $\rho_E=\mathrm{20\,erg\,cm^{-3}}$ (measured in the blob's frame of reference) to reproduce the flux observed by MAGIC and \textit{Fermi}-LAT (see the bottom panel of Fig.~\ref{fig5}). 
Note that such large values of the Lorentz factor of the emission region in the jet have been already postulated in terms of other models in order to explain extremely short flares observed in this source (\citealp{ac14}) or in the other sources, e.g. PKS\,2155-304 (\citealp{ah07}).
Such large Lorentz factors of the blob find also some observational support from the observations of the superluminal motion in PKS\,1510-089 which represents similar type of blazar (\citealp{jo05}).
We can estimate the power of the blob in the observer's reference frame on, 
$L_{\rm blob} = \pi R_{\rm b}^2c\rho_E \gamma_{\rm b}^2\approx 1.7\times 10^{45}$~erg~s$^{-1}$. 
On the other hand, the Eddington luminosity of the black hole in PKS\,1222+21, with the mass $6-8\times 10^8$ M$_\odot$, is $L_{\rm Edd} = 1.3\times 10^{47}M_9\approx 8-10\times 10^{46}$~erg~s$^{-1}$. Therefore, the blob has to contain about $\sim$2$\%$ of the Eddington power. This is quite demanding but seems not to be excluded, especially if $R_{\rm b} \approx R_{\bot}$.  
Lower values of $\gamma_{\rm b}$ require a much higher energy density in the blob (e.g. $\rho_E=\mathrm{340\,erg\,cm^{-3}}$ for $\gamma_{\rm b}=50$ and $R_{\rm b}=3\times 10^{14}$\,cm). 
The strong dependence on the $\gamma_{\rm b}$ is a combined result of the transformation of the energy density to the reference frame of the observer and the beaming of the emission in a narrower cone. 
Note that a larger radius of the blob will lower the energy density constrain, e.g. for $R_{\rm b}=10^{15}$\,cm we obtain $\rho_E=\mathrm{4.9\,erg\,cm^{-3}}$ for $\gamma_{\rm b}=100$ and $\rho_E=\mathrm{80\,erg\,cm^{-3}}$ for $\gamma_{\rm b}=50$, at the assumption that there is no competing energy loss process of the electrons at such a large distance from the star.

\section{Conclusion}\label{sec5}

We propose that blobs of relativistic plasma, moving in jets of AGNs, can encounter from time to time luminous stars which form a quasi-spherical halo around the central super-massive black hole. 
The transit time of a star through a conical jet can be estimated as 
$T_{\rm transit} = R_{\bot}/v_\star\approx 0.45\gamma_{30}^{-1}l_{-1}^{3/2}/M_9^{1/2}$~yr,
where the velocity of the star is 
$v_\star = \sqrt{GM_{\rm BH}/l}\approx 7\times 10^8 (M_9/l_{-1})^{1/2}$ cm s$^{-1}$, 
$M_{\rm BH} = 10^9M_9$ M$_\odot$ is the black hole mass in units of the Solar mass, and $G$ is the gravitational constant. 
%In the case of a parabolic, M87-like, jet the transit time is given by 
%$T_{\rm transit}\approx 2l_{-1}/M_9^{1/2}$ yrs.
%Note that perpendicular extends of conical and parabolic jets at parsec scale distances from the jet base are similar. 
\kom{At parsec scale distance from the jet base the jet transit time is similar for a conical and a parabolic, M87-like, jet.}
Therefore, we do not expect significant effects of the jet shape on our final results.
The star passing the inner jet at parsec distance from the SMBH can be responsible for a sequence of outbursts as different blobs reach its position. 
Such high activity period can last from months to years. 
In fact,  a single blob may be also responsible for the observed sequence of flares (e.g. as observed in Mrk 421, \citealp{bu96} or PKS\,2155-304, \citealp{ah07}), if it meets on its path a few stars within a jet.

Luminous stars are also characterised by stellar winds which pressure can balance the pressure of the blob plasma. As a result a shock structure appears around the star. 
The shock radius around the star in a conical jet can be determined from (see e.g. \citealp{bp97}),
$R_{\rm sh}^\star\approx  1.9\times 10^{12}(M_{-5}v_3)^{1/2}\gamma_{30}^{-1}l_{-1}/L_{46}^{1/2}$ cm, 
where $\dot{M} = 10^{-5}M_{-5}M_\odot$ and $v = 10^3v_3$ km s$^{-1}$ are the mass loss rate and the velocity of the stellar wind, and $L_{\rm b} = 10^{46}L_{46}$ erg s$^{-1}$ is the power of the blob in the observer's frame.
Since the surface area of this shock, $\sim \pi R_{\rm sh}^2$, is typically much smaller than the surface area of the considered blob, the presence of the shock should not essentially influence presented above calculations of the $\gamma$-ray emission from the blob.

Our model predicts the appearance of flares which emission is limited mostly to the $\gamma$-ray energy range. 
Therefore, it can provide the mechanism for the appearance of the so-called orphan $\gamma$-ray flares observed occasionally from relatively nearby AGNs of the BL Lac type (e.g.~ 1ES 1959+650, \citealp{kr04} or Mrk 421, \citealp{fr15}) but also from the FSRQs (e.g. PKS\,1222+21, \citealp{al11}).
In order to show applicability of this scenario we model the case of the emission from the very strong $\gamma$-ray flare observed from the distant FSRQ PKS\,1222+21 in 2010. 
Our model is able to reproduce the observed light curve and spectral energy distribution measured by the {\it Fermi}-LAT and MAGIC, although the energy requirements are quite demanding. Note that in the case of blazars with orphan flares which are about an order of magnitude closer to the observer (i.e. 1ES 1959+650 and Mrk 421), this energy requirements should be about two orders of magnitude lower or the Lorentz factors of the blob 
might be closer to values typically reported from radio observations.

\section*{Acknowledgements}
We thank the Referee for useful comments.
This work is supported by the grant through the Polish Narodowe Centrum Nauki No. 2014/15/B/ST9/04043.

%%%%%%%%%%%%%%%%%%%%%%%%%%%%%%%%%%%%

% Don't change these lines
\bsp	% typesetting comment
\label{lastpage}
\end{document}